\documentclass{XrU2005}
\usepackage{epsfig,graphicx}

\title{Supergiant Fast X-ray Transients: A new class of high mass X-ray
  binaries unveiled by {\it INTEGRAL}} 
\author[1]{I.~Negueruela}
\affil[1]{Dpto. de F\'{\i}sica, Ing. de Sistemas y Teor\'{\i}a de 
la Se\~{n}al, Universidad de Alicante, Apdo. 99, E03080 Alicante, Spain}
\author[2]{D.~M. Smith}
\affil[2]{Physics Dept \& Santa Cruz Institute for Particle
  Physics, University of California Santa Cruz, 1156 High St., Santa
  Cruz, CA 95064}
\author[3]{P.~Reig}
\affil[3]{IESL (FORTH) \& Physics Dept.,University of Crete,  PO Box
  2208, 71003 Heraklion, Crete, Greece}  
\author[4]{S.~Chaty}
\affil{AIM (UMR 7158 CEA/CNRS/Universit\'e Paris 7 Denis Diderot), CEA
  Saclay, DSM/DAPNIA/Service d'Astrophysique, B\^at. 709, 
  L'Orme des Merisiers, 91191 Gif-sur-Yvette, Cedex, France}
\author[1]{J.~M.~Torrej\'on}

\def\la{\mathrel{\mathchoice {\vcenter{\offinterlineskip\halign{\hfil
$\displaystyle##$\hfil\cr<\cr\sim\cr}}}
{\vcenter{\offinterlineskip\halign{\hfil$\textstyle##$\hfil\cr
<\cr\sim\cr}}}
{\vcenter{\offinterlineskip\halign{\hfil$\scriptstyle##$\hfil\cr
<\cr\sim\cr}}}
{\vcenter{\offinterlineskip\halign{\hfil$\scriptscriptstyle##$\hfil\cr
<\cr\sim\cr}}}}}

\begin{document}


\maketitle

\begin{abstract}
{\it INTEGRAL} monitoring of the Galactic Plane is revealing a growing
number of recurrent X-ray transients, characterised by short outbursts
with very fast rise times ($\sim$ tens of minutes) and typical
durations of a few hours. Here we show that several of these transients
are associated with OB supergiants and hence define a new class of
massive X-ray binaries which we call Supergiant Fast X-ray
Transients. Many other transient X-ray sources display similar X-ray
characteristics, suggesting that they belong to the same class. Since
they are difficult to detect and their number is growing fast and
steadily, they could represent a major class of X-ray binaries. 
\end{abstract}

\section{Introduction}

High Mass X-ray binaries (HMXBs) are X-ray sources composed of an
early-type massive star and an accreting compact object. Most known
HMXBs are Be/X-ray binaries, systems consisting of a neutron star
accreting from the disc around a Be star. Even 
though a few Be/X-ray binaries are persistent weak X-ray sources (with
$L_{{\rm X}}\sim10^{34}\:{\rm erg}\,{\rm s}^{-1}$), the majority are
transients, displaying bright outbursts with typical duration on the
order of several weeks.

The second major class of HMXBs contains early-type supergiants. The
compact object is fed by accretion from the strong radiative wind of
the supergiant. These objects are persistent sources, with
luminosities around $L_{{\rm X}}\sim10^{36}\:{\rm erg}\,{\rm s}^{-1}$,
very variable on short timescales, but rather stable in the long
run. Because of their relative brightness and persistent nature, it
has been generally assumed that Supergiant X-ray Binaries (SGXBs) were
easy to detect. About a dozen SGXBs were known before the launch of
{\it INTEGRAL}, most of them having been discovered in the early days of X-ray
astronomy. This low number was generally attributed to a real scarcity
of such systems, as the short duration of the supergiant phase would
result in very short lifetimes.

Since the launch of {\it INTEGRAL}, the situation is changing
dramatically. Several new sources have been detected displaying the
typical characteristics of SGXBs (Walter, these proceedings). In most
cases, the sources had not been detected by previous missions due to
high absorption, which renders their spectra very hard. Here we
show that an even larger population of X-ray sources with OB
supergiant companions may lie hidden in the Galaxy, undetected because
of its transient nature.

\section{X-ray sources with fast outbursts}

{\bf XTE~J1739$-$302 = IGR~J17391$-$3021}

XTE~J1739$-$302 was discovered during an outburst in August 1997
\citep{smi98}. Further observations, mostly with {\it RossiXTE}, but
also with {\it ASCA} showed it to be a strange transient with very
short outbursts, lasting only a few hours \citep{smi05}. Monitoring of
the Galactic Centre region with {\it INTEGRAL} reveals that flares are
rare, with typical intervals between outbursts
of several months \citep{sgue05}. 

The outbursts start with a very sharp rise (with a timescale $<1$ h)
and sometimes show complex structure, with several flare-like peaks
\citep{lut05a,sgue05}. The X-ray spectrum during the outbursts is
generally very absorbed, though the absorption is variable. Good fits
can be achieved with either a power law with a high-energy cut-off or
a thermal bremsstrahlung model with $kT\sim 20\:$keV
\citep{smi05,lut05a}. Such spectra are typical of accreting neutron stars in
a HMXB. The luminosity at the peak of the outbursts approaches $L_{{\rm
     X}}\sim10^{36}\:{\rm erg}\,{\rm s}^{-1}$, also typical of HMXBs. 

 \begin{figure}
\centering
 \begin{picture}(240,220)
\put(0,0){\includegraphics{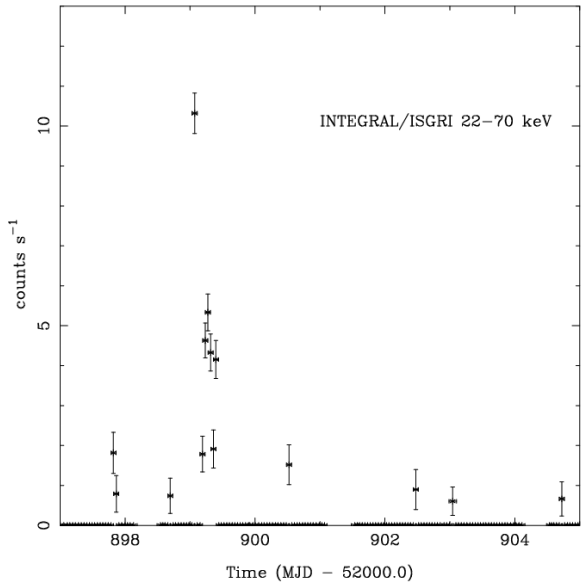}}
\end{picture}
\caption{A typical outburst from a SFXT. {\it INTEGRAL} lightcurve for
IGR~J17544$-$2619 during the flare on 2003 September 17th. The data
have been downloaded from the public data archive at the ISDC.}
\label{fig:outburst}
\end{figure}

The source was not detected during most of an {\it ASCA} pointing in
March 1999 (with an upper
limit $L_{{\rm X}}<10^{33}\:{\rm erg}\,{\rm s}^{-1}$), but went into
outburst at the end of the same observation. 
{\it Chandra} detected the source at a moderate luminosity $L_{{\rm
     X}}\sim10^{34}\:{\rm erg}\,{\rm s}^{-1}$, allowing the identification
     of the counterpart \citep{smi05}. VLT/FORS1 spectra taken in May
     2004 show the 
     counterpart to be an O8\,Iab(f) star, placed at a distance
     $\approx 2.6\:$kpc \citep{neg05}. Interstellar absorption is much
     lower than the absorption implied by X-ray spectral fits.

{\bf IGR~J17544$-$2619}

IGR~J17544$-$2619 was discovered by {\it INTEGRAL} during a short
flare ($\sim 2\:$h; see Fig.~\ref{fig:outburst}) on 2003 September
17th \citep{sun03}. 
Six hours later, it showed a longer (8 h) double-peaked outburst
\citep{greb03}. On 2004 March 8th, it showed a 
complex outburst lasting more than $8\:$h \citep{greb04}.

The source was observed by {\it XMM-Newton} on 2003 September 11th and
17th and in both cases seen at $L_{{\rm X}}\sim10^{35}\:{\rm erg}\,{\rm
  s}^{-1}$ \citep{gonz04}, though it was not detected during a
serendipitous observation in March 2003
($L_{{\rm X}}\la2\times10^{32}\:{\rm erg}\,{\rm s}^{-1}$). {\it
  Chandra} may have observed its quiescent state on 2004 July 3rd, as
it was detected at only $L_{{\rm X}}\sim5\times10^{32}\:{\rm erg}\,{\rm
  s}^{-1}$  and displaying a soft spectrum \citep{zand05}. In
outburst, the spectrum is hard and moderately absorbed, with evidence
for some variation in the amount of absorbing material.

\begin{figure}
\centering
 \begin{picture}(240,180)
\put(0,0){\includegraphics{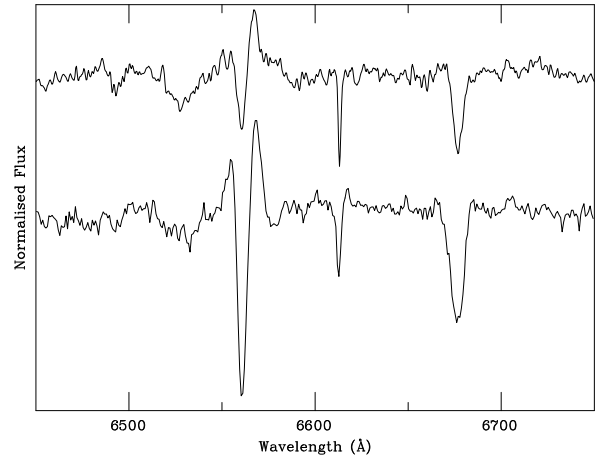}}
\end{picture}
\caption{H$\alpha$ spectra of the counterparts to IGR~J17544$-$2619
  (top) and IGR~J16465$-$4507 (bottom). Both display weak P-Cygni
  profiles, indicative of moderate mass loss.}
\label{fig:two}
\end{figure}

The counterpart to the source has been unambiguously identified with
the {\it XMM-Newton} and {\it Chandra} positions. Spectra taken with
NTT/EMMI show it to be an O9\,Ib supergiant, with a weak wind (see
Fig.~\ref{fig:two}) at a distance of $\sim 3\:$kpc \citep{pel06}.

{\bf IGR J16465$-$4507}

IGR J16465$-$4507 was discovered by {\it INTEGRAL} during an
X-ray flare on 2004 September 7th \citep{lut04}.  A subsequent {\it
  XMM-Newton} observation \citep{lut05b} revealed that the source is a pulsar 
with $P_{{\rm spin}}= 228\:{\rm s}$ and is extremely absorbed ($N_{{\rm
      H}} \sim7\times10^{23}\:{\rm cm}^{-2}$).
 
\begin{figure*}
\centering
 \begin{picture}(480,260)
\put(0,0){\includegraphics{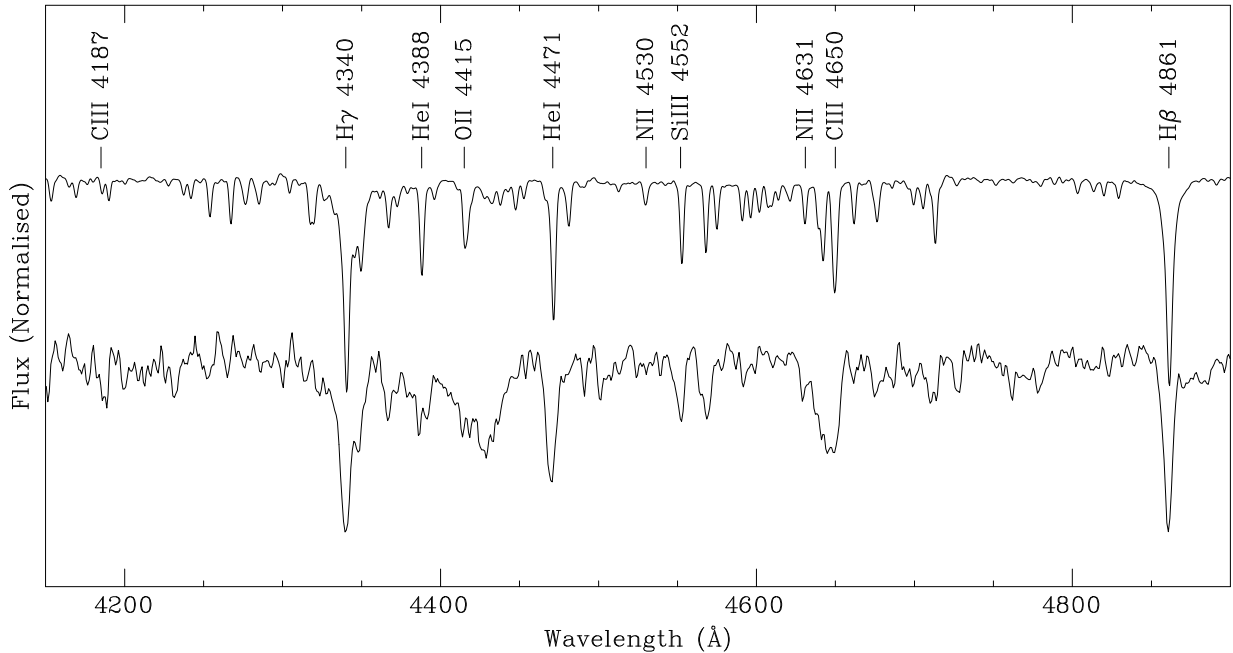}}
\end{picture}
\caption{The spectrum of the optical counterpart to IGR~J16465$-$4507
  (bottom), compared to that of the B1\,Ib supergiant $\zeta$~Per,
  rebinned to the same dispersion. The strength of the metallic lines
  indicates that the counterpart is a supergiant.}
\label{fig:bsg}
\end{figure*}

No further flares have been reported, but the {\it XMM-Newton} error
circle contains only one star. NTT/EMMI spectra of this object were
taken in February and March 2005. The blue spectrum, displayed in
Fig.~\ref{fig:bsg}, is rather noisy, but allows an approximate
classification. While the object is very obviously an early B-type
star, the strength of all metallic lines indicates that it is a
supergiant. Comparison to the spectrum of the B1\,Ib supergiant $\zeta$~Per
  rebinned to the same dispersion clearly shows that the lines are
  very broad for a supergiant, suggesting a very high rotational
  velocity, a typical characteristic of HMXBs.

The presence of strong \ion{C}{iii}~4650\AA\ and moderate
\ion{Si}{iii} lines, while 
\ion{He}{ii}~4686\AA\ is absent, is only compatible with a luminous
star in the B0$-$B1 range. Unfortunately, the signal to noise of the
spectrum falls
below $\sim 20$ around $\lambda$4100\AA\ and we cannot determine the
strength of the \ion{Si}{iv} lines that would allow an exact
classification. The H$\alpha$ spectrum (see Fig.~\ref{fig:two}) shows
evidence for a moderate mass loss. 

{\bf AX 1845.0$-$0433}

AX~1845.0$-$0433 was discovered by {\it ASCA} during a strong flare in
1993. The outburst consisted of a very fast rise (on the order of a
few minutes) followed by a number of peaks during the next few
hours. The spectrum was well fit by an absorbed power law
\citep{yam95}. No further X-ray activity has been reported.

\begin{figure}
\centering
 \begin{picture}(280,280)
\put(0,0){\includegraphics{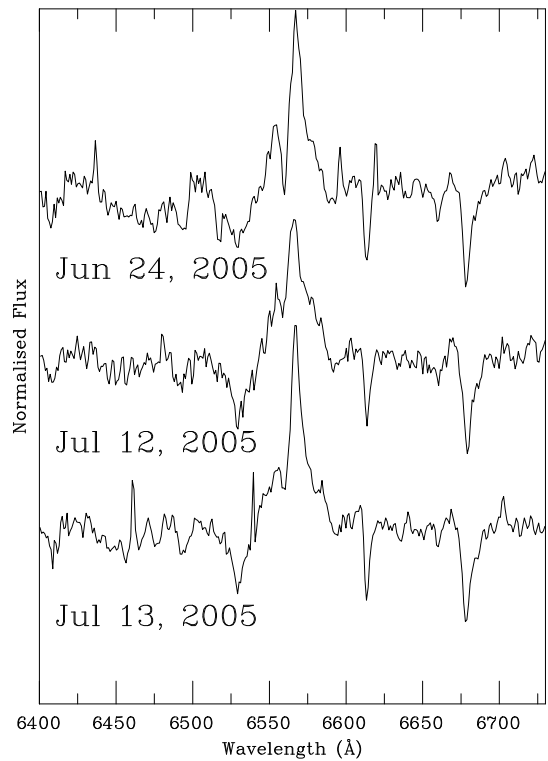}}
\end{picture}
\caption{H$\alpha$ spectra of the proposed counterpart to
 AX~1845.0$-$0433, displaying strong variability, even between
 consecutive nights, a typical signature of an interacting binary.}
\label{fig:axalpha}
\end{figure}

The {\it ASCA} error circle was studied by \citet{coe96}, who found
only one remarkable object, a late O-type supergiant. This star has
been monitored with the 1.3-m telescope at Skinakas observatory. Some
of the spectra are shown in Fig.~\ref{fig:axalpha}. The star shows
strong $H\alpha$ emission, typical of a luminous supergiant.
 Both the shape and strength of the line are variable, sometimes from
 night to night. Such variability is a typical signature of an
 interacting binary. 

{\bf AX J1841.0$-$0536}

AX~J1841.0$-$0536 was observed as a violently variable transient by
{\it ASCA} in April 1994 and then again in October 1999
\citep{bam01}. The source showed multi-peaked flares with a sharp rise
(tenfold increase in count-rate over $\sim1\:$h). 
Analysis of the {\it ASCA} data revealed that the source is a pulsar 
with $P_{{\rm spin}}= 4.7\:{\rm s}$. The spectrum can be fit by an
  absorbed power law plus iron line \citep{bam01}. Based on the
  detection of X-ray flares with a sharp rise,  \citet{bam01}
  suggested that AX~1845.0$ -$0433, AX J1841.0$-$0536 and XTE
  J1739$-$302 could be members of a class with common physical
  features. 

\begin{figure}
\centering
 \begin{picture}(280,230)
\put(0,0){\includegraphics{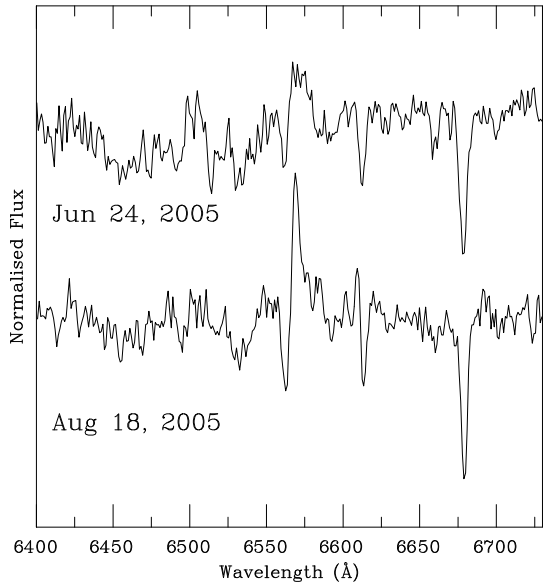}}
\end{picture}
\caption{H$\alpha$ spectra of the counterpart to AX~J1841.0$-$0536,
  showing P-Cygni profiles indicative of the stellar wind of a luminous star.}
\label{fig:axjalpha}
\end{figure}

A {\it Chandra} observation of the field revealed the counterpart to
AX~J1841.0$-$0536, a reddened star with H$\alpha$ in emission
\citep{hal04}. We have taken spectra of this star with the 2.2-m at
Calar Alto and the 1.9-m at SAAO. Though the spectra are very noisy,
they strongly resemble that of the counterpart to IGR~J16465$-$4507, 
 displaying a very strong \ion{C}{iii}~4650\AA\ line and no
\ion{He}{ii}~4686\AA. Therefore, it is also likely to be a luminous
star in the B0$-$1 range.

Some red spectra have been taken with the 1.3-m telescope at Skinakas
observatory (see Fig.~\ref{fig:axjalpha}). Again we find a shape
reminiscent of a P-Cygni profile and strong variability, lending
support to the idea that the object is a supergiant rather than a Be
star. A fast outburst observed by {\it
  INTEGRAL} (IGR~J18410$-$0535) has been attributed to this source
\citep{hg04}. 

\section{Supergiant Fast X-ray Transients: A new class?}
The five sources described in the previous section are characterised
by the occurrence of X-ray outbursts of a very different nature from
those seen in other X-ray binaries. These outbursts are very short
(lasting from $\sim3$ to $\sim8$ hours) and present very sharp rises,
reaching the peak of the flare in $\la 1\:$h. The decay is generally
characterised by a complex structure, with two or three further
flares. 
Three of the sources are associated with O-type supergiants. Though no
pulsations have been detected, they display spectra typical of
accreting neutron stars. The other two sources are X-ray pulsars and
are associated with luminous early B-type stars. In all cases, the
spectra show moderate or high absorption. In the case of
XTE~J1739$-$302 and IGR~J17544$-$2619, which are better studied, the
amount of absorbing material is variable.

We therefore propose that all these objects form a class of HMXBs
which we call Supergiant Fast X-ray Transients (SFXTs), because of the
fast outbursts and supergiant companions. They differ from classical
wind-fed SGXBs, whose X-ray luminosity is variable but always
detectable around $L_{{\rm X}}\sim10^{36}\:{\rm erg}\,{\rm
  s}^{-1}$. Quiescent fluxes of SFXTs
have been near the sensitivity limit of focusing observatories,
with values or upper limits in the range of  $\sim10^{32}$ to
$\sim10^{33}\:{\rm erg}\,{\rm s}^{-1}$.

In spite of this difference, it must be noted that the commonalities
between SFXTs and SGXBs are strong. As a matter of fact, at least
three classical SGXBs have been observed to undergo bright flares on
the same timescale: Vela X-1 \citep{lau95,kri03}, 1E~1145.1$-$6141
\citep{bod04} and Cyg X-1 \citep[][and references
  therein]{gol03}. 


\section{Other sources displaying fast outbursts}
\label{sec:others}

{\bf SAX J1818.6$-$1703}

SAX~J1818.6$-$1703 was discovered by {\it BeppoSAX} during a strong
short outburst (with a rise time of $\sim 1\:$h), in March 1998
\citep{zand98}.  {\it INTEGRAL} detected a double-peaked outburst
in September 2003 \citep{gs05} and
two more in October 2003 \citep{sgue05}. Other fast outbursts have
been observed with the ASM on {\it RossiXTE} \citep{sgue05}.

 \begin{figure}
\centering
 \begin{picture}(280,130)
\put(0,0){\includegraphics{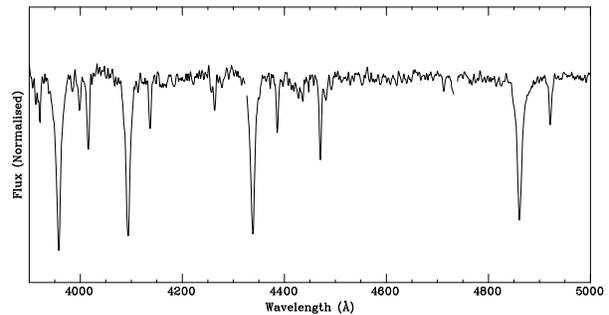}}
\end{picture}
\caption{Optical spectrum of HD 168078.}
\label{fig:nosaxj}
\end{figure}

The X-ray lightcurve of SAX~J1818.6$-$1703 is typical of a SFXT. The
X-ray spectrum is very hard \citep{gs05}. The optical counterpart to
SAX~J1818.6$-$1703 is not known. A bright early type star, HD~168078, is
within both the {\it BeppoSAX} and {\it INTEGRAL} error
circles. Spectra of this object, taken  with the 1.3-m at Skinakas and the
1.9-m at SAAO show a normal B3\,IV star, without indications of
emission (see Fig.~\ref{fig:nosaxj}). As HD~168078 is 
not a convincing counterpart to SAX~J1818.6$-$1703, it might be
worthwhile to search the error circle for a reddened supergiant.

{\bf IGR J16479$-$4514}

IGR~J16479$-$4514 was discovered by {\it INTEGRAL} during an outburst
in August 2003 \citep{mol03}. Several other short outbursts with very
fast rise times were observed by {\it INTEGRAL} during 2003
\citep{sgue05}. The X-ray spectrum of IGR~J16479$-$4514 is a power-law
with high-energy cut-off, typical of
a HMXB with a neutron star \citep{lut05b}. All these characteristics
are strongly suggestive of a SFXT.

{\bf XTE J1901+014}

XTE J1901+014 was discovered by {\it RXTE} during a very bright
outburst (reaching 0.9 Crab) that lasted less than $8\:$h,  in 2002
\citep{rs02}. Analysis of {\it RXTE}/ASM data revealed another fast
outburst in 1997 \citep{rs02}. Several outbursts have been observed by {\it
  INTEGRAL}. The {\it INTEGRAL} error
circle contains two {\it ROSAT} sources and one {\it Einstein} source,
one of which could represent the quiescent state of XTE~J1901+014
\citep{ste05}.   

{\bf AX~J1749.1$-$2733}

AX~J1749.1$-$2733 was observed by {\it ASCA} at a
relatively low luminosity on several occasions \citep{sak02}. It has
recently been reported to show fast 
X-ray outbursts during {\it INTEGRAL} observations (Grebenev, quoted
in In't Zand 2005).

\section{Discussion \& Conclusions}

Most X-ray transients, whether Be/X-ray binaries or low mass X-ray
binaries (having either neutron star or black hole companions), display
outbursts typically lasting from a few weeks to months. Such durations
are compatible with viscous timescales in a typical accretion
disc. Fast outbursts, with much shorter durations, must be due to a
completely different physical mechanism. Here we have shown that at
least a fraction of the recurrent fast X-ray transients are associated
with luminous OB stars, suggesting that the mass transfer mechanism
feeding the accreting compact object is a radiative wind, and therefore
have identified the class of SFXTs.

We must note that not every source displaying short X-ray outbursts
may be assigned to the class of SFXTs. Flare stars and RS CVn binaries
display short X-ray flares and superbursts
from low mass X-ray binaries have similar durations, though rather
different luminosities and lightcurves \citep[e.g.][]{gs05}. The HMXB
1A\,0535$-$668, in the LMC, has shown very bright X-ray outbursts
lasting only a few days, but these outbursts are locked in phase with
the orbital period of the system and may be related to periastron
passage in a very eccentric orbit
\citep[cf.][]{cha83}. IGR~J00370+6122 could be a similar system
\citep{rei05}. 
The black hole transient V4641~Sgr should not be grouped with SFXTs
either. In this system, the mass donor is a B9\,III giant
\citep{oro01}, which is not a massive star and cannot have a radiative
wind. Moreover, its 1999 outburst was highly super-Eddington and
accompanied by a huge optical brightening, properties that set it
completely apart from SFXTs \citep{rev02}.

However, there are strong reasons to believe that the class of SFXTs
comprises a much larger number of sources than the five objects
described above. In
Section~\ref{sec:others}, we list four reliable candidates, but
other known sources might belong to this class. For example, the 
 X-ray transient IGR~J11215$-$5952 was observed by {\it INTEGRAL}
 during the decaying phase of an outburst. It is likely to be
 related to SFXTs if its association to the B1\,Ia supergiant 
HD~306414 is confirmed \citep[cf.][]{atel}.

If only some of these candidates are confirmed, the number of SFXTs
would already be comparable to that of classical SGXBs. We must
consider, however, that classical SGXBs are persistent
bright X-ray sources, while SFXTs are transient sources with very
short duty cycles. Most of them are detectable, unless directly
pointed at by {\it Chandra} or {\it XMM-Newton}, only for a few hours
every several months. It is hence not surprising that most of the
SFXTs so far found lie on the vicinity of the Galactic Centre, a
region extensively monitored by {\it INTEGRAL} and other satellites.

But, if SFXTs are very difficult to detect, and we already know
several of them in the region around the Galactic Centre, the
implication is that the population of SFXTs in the Galaxy is much
larger than the ten or so objects already known. As a matter of
fact, it is difficult to avoid concluding that most binaries
containing a supergiant and a compact object {\bf must} be SFXTs or
entirely quiescent. If we
take into account the large number of obscured persistent HMXBs that 
 {\it INTEGRAL} is discovering, it seems clear that the numbers of
 HMXBs must have
been severely underestimated.

The physical reason for fast outbursts is still unknown. \citet{gol03}
speculated that the outbursts in Cyg X-1 could be due to some form of
discrete mass ejection from the supergiant donor. \citet{zand05} also
suspects that wind variability is the cause of fast outbursts. As we
have shown here, an important fraction of fast transients have
supergiant companions and at least three classical SGXBs have shown
fast outbursts. It seems then that the fast outbursts are related to
the mass transfer mode, wind accretion. They cannot be related to the
nature of the companion, as they are seen in black hole systems (Cyg
X-1), slow X-ray pulsars (IGR J16465$-$4507) and faster X-ray pulsars
(AX J1841.0$-$0536). 

There is increasing evidence suggesting that the winds of B-type
supergiants are highly structured and may have a fundamentally clumpy
nature \citep[][and references therein]{pri05}. If these clumps
survive to the distance at which the compact object is orbiting, they
could give rise to sudden episodes of increased accretion
rate. Alternatively, the outbursts could be related to the instability
believed to be intrinsic to the wind accretion process \citep[e.g.][]{fog05}.

On the other hand, there is nothing in the optical properties of SFXTs
setting them apart from classical SGXBs. It is therefore difficult to
understand why their quiescent X-ray luminosities are rather lower. A
possibility would be that SFXTs have wider orbits than SGXBs, and the
compact object (in most cases, a neutron star) accretes from a less
dense environment. This, however, would not explain why the sources
spend some (still not quantified) fraction of time below
detectability. If highly eccentric, wide orbits are invoked in order
to explain the periods of very low X-ray luminosity, one would
na\"{\i}vely expect some (quasi-)periodicity in the recurrence of the
outbursts, that has not been observed, as they would have to occur
always relatively close to periastron.

Clearly, a more complete investigation of all the sources presented
here is needed before common trends start to emerge and a
characterisation of the group of SFXTs can be achieved. Understanding
the reasons for their X-ray behaviour and the source of the difference
with classical SGXBs will undoubtedly increase our knowledge of the
accretion process and very likely provide valuable insights into the
different paths leading to the formation of HMXBs and their
subsequent evolution.

\section*{Acknowledgments}

IN is a researcher of the
programme {\em Ram\'on y Cajal}, funded by the Spanish MCyT
and the University of Alicante.
This research is partially supported by the Spanish MCyT under grants
AYA2002-00814 and ESP-2002-04124-C03-03.

Partly based on observations collected at the European
  Southern Observatory, La Silla, Chile (ESO 73.D-0081;
  274.D-5010) and observations collected at the
  DSAZ, Calar Alto, operated by
  the MPI f\"ur Astronomie Heidelberg jointly with the
  Spanish CNA. We thank Calar Alto for
  allocation of director's discretionary time to this
  programme. Skinakas Observatory is a collaborative project of the
  University of Crete,  
the Foundation for Research and Technology-Hellas and the
  MPI f\"ur Extraterrestrische Physik.  This  
  research has made use of the Simbad data base, operated at CDS,
Strasbourg, France.    
 
We are very grateful to Jorge~Casares for carrying out the
observations at SAAO. We also thank Leonardo Pellizza for advancing
results on IGR~J17544$-$2619.

\end{document}